
%
\magnification=\magstep1
\baselineskip=13pt
\overfullrule=0pt
\nopagenumbers
\font\twelvebf=cmbx12

\rightline{UdeM-LPN-TH-94-191}
\rightline{CRM-2163}
\rightline{solv-int/9402004}
\rightline{Feb 1994}
\vskip .3in
\centerline{\twelvebf On UrKdV and UrKP}
\vskip .3in
\centerline{Didier A. Depireux}
\smallskip
\centerline{\it Laboratoire de Physique Subnucl\'eaire and}
\centerline{\it Centre de Recherches Math\'ematiques,}
\centerline{\it Universit\'e de Montr\'eal, C.P.6128, succ. Centre-Ville}
\centerline{\it Montr\'eal, Canada H3C 3J7}
\smallskip
\centerline{and}
\smallskip
\centerline{Jeremy Schiff}
\smallskip
\centerline{\it Department of Mathematics and Computer Science}
\centerline{\it Bar Ilan University}
\centerline{\it Ramat Gan 52900, Israel}
\vskip .5in
\centerline{\bf Abstract}
\smallskip
\noindent We present two extensions of Wilson's explanation of the
Miura map from MKdV to KdV. In the first we explain the map of
Svinolupov {\it et.al.} from a certain UrKdV-like equation to KdV,
and in the second we explain Konopelchenko's map from the modified
KP equation to KP. In the course of the latter we introduce an
``UrKP'' system, with an infinite dimensional symmetry, providing
us with a systematic method to construct B\"acklund transformations
for the modified KP and KP equations.
\vskip .3in
\centerline{\bf R\'esum\'e}
\noindent Nous donnons deux g\'en\'eralisations du travail pr\'esent\'e
par Wilson sur l'explication de la carte de Miura de l'\'equation de
mKdV \`a celle de KdV. La premi\`ere g\'en\'eralisation justifie la
carte donn\'ee par Svinolupov {\it et.al.} d'une certaine \'equation
reli\'ee \`a l'\'equation de UrKdV \`a celle de KdV elle-m\^eme.
Ensuite, comme seconde g\'en\'eralisation, nous expliquons la carte
present\'ee par Konopelchenko sur le passage de l'\'equation de KP
modifi\'ee \`a celle de KP. Dans cette seconde partie, nous introduisons
un syst\`eme que nous baptisons ``UrKP'', qui poss\`ede une sym\'etrie
de dimension infinie et qui nous fournit une m\'ethode syst\'ematique
pour d\'eriver les transformations de B\"acklund pour les \'equations de
KP modifi\'ee et de KP.
\noindent
\vfill\eject

\baselineskip=15pt
\footline={\hss\tenrm\folio\hss}
\pageno=1

\noindent{\bf 1.Introduction}

\smallskip

Probably the most important insight in the theory of integrable
systems in recent years is Wilson's explanation of the Miura map [1].
If $j$ satisfies the MKdV equation
$$ j_t={\textstyle{1\over 4}}j_{xxx}-{\textstyle{3\over 8}}j^2j_x~,
           \eqno{(1)}$$
then $u=-{1\over 2}(j_x-{1\over 2}j^2)$ solves the KdV equation
$$ u_t={\textstyle{1\over 4}}u_{xxx}-{\textstyle{3\over 2}}uu_x~.
            \eqno{(2)}$$
Now in general if $v$ is some function of $j$ and its $x$-derivatives,
we can compute $v_t$ using (1), but we should not expect to be able
to write $v_t$ as a function of $v$ and its $x$-derivatives.
As Wilson explains, the fact that
$u_t$ {\it can} be written in terms of $u$ and its $x$-derivatives
suggests strongly that (1) has some symmetry, and that $u$ is
invariant under this symmetry. To see that this is indeed the
case it is necessary to introduce the UrKdV equation
$$ q_t={\textstyle{1\over 4}}\left(q_{xxx}-{{3q_{xx}^2}\over {2q_x}}
          \right)~.   \eqno{(3)}$$
Via the map $j=q_{xx}/q_x$, a solution of UrKdV
generates a solution of MKdV, which in turn generates, via the Miura
map, a solution $u=-{1\over 2}(q_{xxx}/q_x-3q_{xx}^2/2q_x^2)$ of
KdV. Now (3) can be seen to be invariant under the group of M\"obius
transformations $q\rightarrow (aq+b)/(cq+d)$, $ad-bc=1$. The reason
$j_t$ can be written in terms of $j$ and its derivatives is that
$j$ is (the in some sense unique) invariant under the $c=0$
subgroup of M\"obius transformations; similarly the reason $u_t$
can be written in terms of $u$ and its derivatives is that $u$ is
(the in some sense unique [2]) invariant under the full
group of M\"obius transformations. The Miura map, which gives $u$ in
terms of $j$, reflects the fact that an invariant of the full group
is of course an invariant of any subgroup; but since the
$c=0$ subgroup of M\"obius transformations is not a normal subgroup
of the full group, the symmetry of MKdV which ``explains'' the Miura
map is not a group symmetry, and cannot be properly explained without
introducing UrKdV.

In [3], one of us noted that $\tilde{j}=q_{xx}/q_x-2q_x/q$ is
(the in some sense unique) invariant under $b=0$ M\"obius
transformations, and $\tilde{j}$ also satisfies MKdV. From
the relation between $j$ and $\tilde{j}$ we deduce the $x$-part of
a (strong) B\"acklund transformation for MKdV; if $j$ satisfies
MKdV, so does $j-2/r$, where $r$ satisfies
$$ \eqalign{r_x+rj&=1~,\cr
            r_t+{\textstyle{1\over 4}}r(j_{xx}-{\textstyle{1\over 2}}
      j^3)&={\textstyle{1\over 4}}(j_x-{\textstyle{1\over 2}}j^2)~. }
         \eqno{(4)}$$
In greater generality, any equation which arises from a group
invariant equation as the equation satisfied by a quantity invariant
under a (proper) subgroup, will display B\"acklund transformations.
This is a powerful way to construct B\"acklund transformations, but
such transformations in general give rise to only rather limited
classes of solutions.

{}From what we have said above, it is apparent that the
the Miura map and the B\"acklund transformation (4) are purely
consequences of the M\"obius invariance of the UrKdV equation (3).
We could have started with any evolution equation of the form
$$ q_t=q_xF(u,u_x,u_{xx},...)~,   \eqno{(5)}$$
and deduced
1) that the Miura map maps
solutions of $$ j_t=\partial_x(\partial_x+j)F  \eqno{(6)}$$
to solutions of
$$ u_t=(-{\textstyle{1\over 2}}\partial_x^3+u\partial_x+\partial_xu)F~,
       \eqno{(7)}$$
and 2) that if $j$ satisfies (6) so does $j-2/r$, where
$$ \eqalign{r_x+rj&=1~,\cr
            r_t+r(F_x+jF)&=F~. } \eqno{(8)}$$
In particular the Miura map and the B\"acklund transformation (4)
of MKdV are {\it not} reflections of the integrability of KdV and
its relatives, as is often argued.

The original aim of the current work was to extend Wilson's ideas
to explain the KP-Miura map [4], namely that a solution of MKP
$$ j_t={\textstyle{1\over 4}}j_{xxx}
     -{\textstyle{3\over 8}}j^2j_x
     +{\textstyle{3\over 4}}(\partial_x^{-1}j_{yy}-j_x\partial_x^{-1}j_{y})
    \eqno{(9)}$$
gives a solution of KP
$$ u_t={\textstyle{1\over 4}}u_{xxx}-{\textstyle{3\over 2}}uu_x
          +{\textstyle{3\over 4}}\partial_x^{-1}u_{yy}   \eqno{(10)}$$
via $u=-{1\over 2}(j_x-{1\over 2}j^2-\partial_x^{-1}j_y)$.
The resulting theory turns out to be quite rich, and we will
give only a part of it here (section 3). But {\it en route} to this
theory we observed that Wilson's ideas actually have an extension for
the KdV system that seems of some importance.
It has been observed [5] that for arbitrary constants $A,B$,
a solution of the equation
$$ \phi_t = {\textstyle{1\over 4}} \left(\phi_{xxx}-{{3\phi_{xx}^2}
     \over{2\phi_x}} + {{3(B^2-4A\phi)}\over{2\phi_x}}
     \right)
     \eqno{(11)}$$
gives a solution of KdV (2) via the map
$$ u =  {\textstyle{1\over 2}} \left({{\phi_{xxx}}\over{\phi_x}}
     -{{\phi_{xx}^2}\over{2\phi_x^2}} + {{B^2-4A\phi}\over{2\phi_x^2}}
     \right)~.
     \eqno{(12)}$$
In section 2 we give the group-theoretical explanation of this
map.  Setting $A=B=0$ we deduce from (12)
that there is a second map from UrKdV
to KdV  in addition to the standard
``Schwartzian derivative'' one given after (3). The existence of these two
maps from UrKdV to KdV is equivalent to the $j\rightarrow -j$
symmetry of MKdV.  It is precisely this
feature of MKdV and the higher equations in the MKdV
hierarchy that make them unique amongst the equations of form (6) [6].

\vskip .2in

\noindent{\bf 2.Extension of the UrKdV formalism}

\smallskip

As is well known, the KdV equation has a zero curvature
formulation:
$$ \partial_tM-\partial_xP+[P,M]=0 ~, \eqno{(13)}$$
where
$$ P=\pmatrix{-{1\over 4}u_x & -{1\over 2} u \cr
             {1\over 4} u_{xx} - {1\over 2}u^2& {1\over 4}u_x \cr}
 \qquad , \qquad
  M=\pmatrix{0&1\cr u&0\cr}~. \eqno{(14)}$$
$P$ and $M$ are $sl(2)$ matrices. The origin of the UrKdV equation
can be understood as follows: (13) is the consistency condition for
the equations
$$ \eqalign{ g_t&=-Pg \cr
             g_x&=-Mg } \eqno{(15)}$$
where $g$ is an $SL(2)$ matrix. The second equation in (15)
essentially
determines all entries in $g$ in terms of one unknown function $q$,
and gives $u$ in terms of this unknown function
$q$; the first equation in (15) gives the
evolution of $q$. But (15) is invariant under $g\rightarrow gh$, where
$h$ is a constant $SL(2)$ matrix; it follows that whatever the
evolution equation for $q$ is, it will have an SL(2) invariance, and
this gives rise to the M\"obius invariance of UrKdV.

Similarly, any equation of form (7) can be written in the form (13),
with
$$ P=\pmatrix{{1\over 2}F_x&F\cr
               Fu-{1\over 2}F_{xx}&-{1\over 2}F_x\cr}
  \qquad , \qquad
   M=\pmatrix{0&1\cr u&0\cr}~,  \eqno{(16)}$$
and thus the origin of the M\"obius invariant equation (5) can be
understood. From the point of view of zero curvature formulations,
the feature that distinguishes the equations of the KdV hierarchy from
other equations of form (7), is that they have a one-parameter family
of zero curvature formulations; for KdV we can take in (13)
$$ P=\pmatrix{-{1\over 4}u_x & \lambda-{1\over 2} u \cr
             {1\over 4} u_{xx} +(\lambda+u)(\lambda- {1\over 2}u)&
             {1\over 4}u_x \cr}
   \qquad , \qquad
  M=\pmatrix{0&1\cr u+\lambda&0\cr}~, \eqno{(17)}$$
for any $\lambda$ ($\lambda$ is usually called the spectral parameter).
(13), with the choice (17),
can be regarded as the consistency condition for equations of
the form (15), but we now  take $g$ as a $2\times 2$ matrix
which is a formal power series in $\lambda$, that is
$$ g=\sum_{n=0}^{\infty} g_n\lambda^n,  \eqno{(18)}$$
where the $g_n$ are $2 \times 2$
matrices\footnote{$^1$}{\baselineskip=22pt We could
restrict $g$ by requiring that $\det(g)$,
which can be computed as a formal power series in $\lambda$, be 1.
We would then only discover the map (12) for $A=1$, $B=0$.
By rescaling and translating $\phi$ one can deduce the map for
arbitrary $A\not=0$ and $B$ from this case, but the case $A=0$ is also
of interest.}.

What is now the content of equations (15)? The first row of the
second equation constrains $g_n$ to be of the form
$$ g_n=\pmatrix{ \alpha_n&\beta_n\cr-\alpha_{nx}&-\beta_{nx}\cr},
              \eqno{(19)}$$
while the second row gives the following relations
$$ \eqalign{\alpha_{0xx}-u{\alpha_0}
              &=\beta_{0xx} -u{\beta_0}=0~, \cr
            \alpha_{n-1}&=\alpha_{nxx}-u\alpha_n~,~~~n>0 \cr
            \beta_{n-1}&=\beta_{nxx}-u\beta_n~,~~~n>0. \cr} \eqno{(20)}$$
For each $N$ we can use the equations (20) for $n\le N$ to write
$\alpha_0,\beta_0,\alpha_1,\beta_1,...,\alpha_{N-1},\beta_{N-1}$ in
terms of $u,\alpha_N,\beta_N$, and we are left with two relations
between the three functions $u,\alpha_N,\beta_N$. Assuming we can solve
these relations and write $u$ in terms of the pair $\alpha_N,\beta_N$,
which satisfy a single constraint, we can then consider the first
equation of (15), which when truncated at order $\lambda^N$ gives a
consistent evolution for the constrained pair $\alpha_N,\beta_N$; this
evolution must induce the KdV evolution (2) for $u$. We will compute
this evolution for $N=1$ shortly\footnote{$^2$}{\baselineskip=22pt For
higher $N$ it seems one can only solve for $u$ in terms of
$\alpha_N,\beta_N$ in a formal sense, and the evolutions are nonlocal.}.
But first we note that equations (15) have an infinite dimensional
invariance, the invariance $g\rightarrow gh$, where $h$ is a matrix
valued formal power series in $\lambda$. Writing
$h=\sum_{n=0}^{\infty}h_n\lambda^n$, we see that if we allow
transformations with $\det(h_0)=0$, we have only a monoid invariance.
For ease, we restrict to those $h$ for which $\det(h_0)\not=0$, to
obtain an infinite dimensional group invariance. The dimension of the
subgroup that acts nontrivially on any particular $g_N$, however, is
finite. This gives the symmetry group of the evolution equation for the
constrained pair $\alpha_N,\beta_N$.

To implement the above procedure for $N=1$, it is useful to observe
that equations (20) imply that
$$ \eqalign{ \partial_x(\alpha_0\beta_{0x}-\alpha_{0x}\beta_{0})&=0~,\cr
 \partial_x(\alpha_{0}\beta_{1x}-\alpha_{0x}\beta_{1}
           +\alpha_{1}\beta_{0x}-\alpha_{1x}\beta_{0})&=0~.\cr}
                   \eqno{(21)}$$
We therefore define
$$ \eqalign{ A&=\alpha_0\beta_{0x}-\alpha_{0x}\beta_{0}~,\cr
             B&=\alpha_{0}\beta_{1x}-\alpha_{0x}\beta_{1}
           +\alpha_{1}\beta_{0x}-\alpha_{1x}\beta_{0}~.\cr}
                   \eqno{(22)}$$
It can easily be checked that the evolutions for $\alpha_0,\beta_0,
\alpha_1,\beta_1$ obtained from (15) imply that $A_t=B_t=0$, and thus
$A$ and $B$ are constant. Having introduced $A$ and $B$,
it is now straightforward to deduce from (20) and (22) that
$$ u={{(\alpha_1\beta_1'''-\beta_1\alpha_1''')-
         (\alpha_1'\beta_1''-\beta_1'\alpha_1'')-B}\over
     {2(\alpha_1\beta_1'-\beta_1\alpha_1')}}
\eqno{(23)}$$
(here primes denote differentiation with respect to $x$),
and that the constraint between $\alpha_1$ and $\beta_1$ is
$$ A=u'(\alpha_1\beta_1''-\beta_1\alpha_1'')
     + u^2(\alpha_1\beta_1'-\beta_1\alpha_1')
     - u(\alpha_1\beta_1'''-\beta_1\alpha_1'''
          + \alpha_1''\beta_1'-\beta_1''\alpha_1')
     + (\alpha_1''\beta_1'''-\beta_1''\alpha_1''').\eqno{(24)}$$
The evolution equations for $\alpha_1,\beta_1$ are
$$ \eqalign{\alpha_{1t}&=\alpha_1'''-{\textstyle{3\over 4}}\alpha_1 u'
                                 -{\textstyle{3\over 2}}\alpha_1' u ~,\cr
            \beta_{1t} &= \beta_1'''-{\textstyle{3\over 4}}\beta_1 u'
                                 -{\textstyle{3\over 2}}\beta_1' u~. \cr
      } \eqno{(25)}$$
One can explicitly check that these flows induce the KdV flow for
$u$, and preserve the constraint (24). For completeness we also
write down the evolutions of $\alpha_0,\beta_0$:
$$ \eqalign{\alpha_{0t}&={\textstyle{1\over 4}}\alpha_0 u'
                                 -{\textstyle{1\over 2}}\alpha_0' u~, \cr
            \beta_{0t} &= {\textstyle{1\over 4}}\beta_0 u'
                                 -{\textstyle{1\over 2}}\beta_0' u~. \cr
      } \eqno{(26)}$$
The symmetry group of (25) is given by
$$ (~\alpha_1~~\beta_1~) \rightarrow
   (~\alpha_0~~\beta_0~)h_1 + (~\alpha_1~~\beta_1~)h_0~.  \eqno{(27)}$$
The group is a semidirect product of the group of invertible $2\times
2$ matrices under matrix multiplication with the group of $2 \times 2$
matrices under matrix addition; elements of
the group are pairs $(h_0,h_1)$ of $2\times 2$ matrices, with the
product
$$ (h_0,h_1).(\tilde{h}_0,\tilde{h}_1)=(h_0\tilde{h}_0,
                 h_0\tilde{h}_1+h_1\tilde{h}_0)~.   \eqno{(28)}$$
Note that $A,B$ transform nontrivially under symmetry group elements:
$A$ transforms nontrivially if and only if $det(h_0)\not=1$;
$B$ transforms nontrivially only if $det(h_0)\not=1$ or
$Tr(h_1h_0^{-1}\not=0$).

Following the philosophy of Wilson, we try to construct invariants
of subgroups of the symmetry group. $\alpha_0,\beta_0$ are invariant
under the normal subgroup $\{(h_0,h_1)~\vert~h_0=I\}$, and (where it
is defined)
$q=\beta_0/\alpha_0$ satisfies the UrKdV equation (displaying a
group symmetry since we have factored out a normal subgroup). The
quantity
$$ \phi=\alpha_1\beta_{1x}-\alpha_{1x}\beta_{1}  \eqno{(29)}$$
is an invariant under the non-normal subgroup
$\{(h_0,h_1)~\vert~\det(h_0)=1, h_1=0\}$ .
It is straightforward but tedious to
check that $\phi$ defined thus satisfies (11), and further that $u$
is given in terms of $\phi$ by (12).
The symmetry underlying the existence of this map
is thus described; in terms of $\phi$ it is of course nonlocal.

We conclude this section with a reference to [7]; in this paper a
map between a certain coset of a loop group and the space of solutions
of a system essentially equivalent to UrKdV is described, and we
expect there should be a similar description for certain solutions
of the system (25). Also in this paper a somewhat different
explanation of the origin of the UrKdV equation is given, based
on the zero curvature formulation of MKdV as opposed to KdV.

\vskip .2in

\noindent{\bf 3.UrKP Formalism}

\smallskip

The KP hierarchy has a variety of zero curvature formulations.
The one we shall use, which is not the most standard, but
might be regarded as the natural extension of  equations
(13),(17) for KdV, can be inferred from [8]:
$$ \partial_tM-\partial_xP+[P,M]=0 ~, \eqno{(30)}$$
$$ P=\pmatrix{-{1\over 4}u_x
              +{3\over 4}\partial_x^{-1}u_{y}&
      \partial_y-{1\over 2} u \cr
   \noalign{\smallskip}
 {1\over 4} u_{xx}+{3\over 4} u_y +(\partial_y+u)(\partial_y- {1\over 2}u)&
             {1\over 4}u_x
      +{3\over 4}\partial_x^{-1}u_{y}\cr}
 \quad , \quad
  M=\pmatrix{0&1\cr
   \noalign{\smallskip}
             u+\partial_y&0\cr}~. $$
Here $P,M$ belong to the algebra of finite order $2\times 2$ matrix valued
linear differential operators in $y$. There is in fact a zero
curvature formulation of KP using the algebra of finite order
$n\times n$ matrix valued linear ordinary differential operators
for any $n$ [9];
{}from each of these one can extract an ``UrKP'' equation. However here
we focus  just on the implications of (30).

(30) can be regarded as a consistency condition for the
system
$$ \eqalign{ g_t&=-Pg \cr
             g_x&=-Mg } \eqno{(31)}$$
$$ g \in{\cal G} = \left\{ g~\vert~
               g=\sum_{n=0}^{\infty}g_n(x,y,t)\partial_y^n,
                       ~ g_n\in{\cal M}_{2,2} \right\} $$
(${\cal M}_{2,2}$ denotes the set of $2\times 2$ matrices;
${\cal G}$ is a set of formal sums);
{\it alternatively} (30) can be regarded as a
consistency condition for the system
$$ \eqalign{ \tilde{g}_t&=\tilde{g}P \cr
             \tilde{g}_x&=\tilde{g}M } \eqno{(32)}$$
$$ \tilde{g}\in\tilde{\cal G} =
     \left\{ \tilde{g}~\vert~
            \tilde{g}=\sum_{n=0}^{\infty}\partial_y^n
             \tilde{g}_n(x,y,t), ~ \tilde{g}_n\in{\cal M}_{2,2} \right\}~.
           $$
In our considerations of the KdV system, we took $g$ in equation (15)
to lie in a set that (if we exclude a small, uninteresting subset) has
a group structure. Thus there was never a need to consider a system
analogous to
(32) as well as the system (15), since if $g$ satisfies (15),
$\tilde{g}=g^{-1}$ satisfies $\tilde{g}_t=\tilde{g}P$,
$\tilde{g}_x=\tilde{g}M$. Here neither ${\cal {G}}$ nor
$\tilde{\cal G}$ have group structure, so these formulations are
distinct. Equations (31) and (32) make sense because there is
a well-defined natural multiplication of $\cal G$
(resp.$\tilde{\cal G}$)
on the left (resp.right) by finite order matrix
valued linear differential operators in $y$.

For our purposes it is important to identify the symmetries
of the systems (31) and (32). One easily establishes that there is
a well-defined natural multiplication of $\cal G$ (resp.
$\tilde{\cal G}$) on the right (resp.left) by
any $g\in{\cal G}$ (resp.$\tilde{g}\in\tilde{\cal G}$) for which
$g_n$ (resp.$\tilde{g}_n$) is polynomial in $y$, for $n=0,1,2,...$.
{}From this we deduce the following symmetry of (31):
$$ g\rightarrow gh \eqno{(33)}$$
$$ h\in{\cal H}=\left\{ h ~\vert~
             h=\sum_{n=0}^{\infty}h_n(y)\partial_y^n,
             ~h_n\in{\cal M}_{2,2},~h_n~{\rm polynomial~in}~y
                \right\}     $$
(Further consideration of (32) will be deferred to [9]).
The set $\cal H$ has a natural ring structure; indeed if $\cal M$
is an arbitrary ring of matrices we can consider the ring of
operators
$$ {\cal H}_{\cal M}=\left\{ h ~\vert~
             h=\sum_{n=0}^{\infty}h_n(y)\partial_y^n,
             ~h_n\in{\cal M},~h_n~{\rm polynomial~in}~y
                \right\}.     \eqno{(34)}$$
For ${\cal M}=U(p)$ the Lie algebra obtained by supplying ${\cal H}_
{\cal M}$ with the commutator bracket is essentially a classical limit
of the $W_{\infty}^p$ algebras studied by Bakas and Kiritsis [10]
and Odake and Sano [11].
Note that any subring of $\cal M$ gives a subring of ${\cal H}_
{\cal M}$. Here we focus our attention on the transformation of
$g_0$ under the action (33); we have
$$ g_0 \rightarrow \sum_{n=0}^{\infty} g_n h_0^{(n)} \eqno{(35)}$$
where $h_0^{(n)}$ denotes the $n$th $y$-derivative of $h_0$; the
sum is finite since $h_0$ is polynomial. We note that an element
of $\cal H$ with any desired $h_0$ (of degree $m$)
can be constructed by setting
$$\eqalign{h&=h_1h_2\cr
           h_1&=\sum_{n=0}^m h_{1n}\partial_y^n~~~~~~(h_{1n}~{\rm
                       independent~of~}y) \cr
           h_2&=y^m\pmatrix{1&0\cr 0&1\cr}; \cr}
        \eqno{(36)}$$
it follows that
the full symmetry action on $g_0$ is generated by the
transformations
$$\eqalign{ g_0&\rightarrow g_0h_{10}~~~~~(h_{10}~{\rm constant})\cr
            g_0&\rightarrow yg_0+g_1. \cr}
        \eqno{(37)}$$
It is maybe unclear whether the action on $g_0$
is an $\cal H$ action or an ${\cal H}_0$ action, where
${\cal H}_0=\{h\in{\cal H}\vert h_n=0,n>0\}$. Because of
the appearance of all the $g_n$ in the transformation law (35),
there is a full $\cal H$ action.

We now consider the evolution of $g_0$ obtained from (31), and
the relationship of the entries of $g_0$ to $u$.
{}From the second equation of (31) we find we can write
$$ g_0=\pmatrix{ \alpha&\beta\cr-\alpha_{x}&-\beta_{x}\cr},
              \eqno{(38)}$$
where $\alpha,\beta$ are related to $u$ via
$$\eqalign{
\alpha_{xx}&=\alpha_y+u\alpha~, \cr
\beta_{xx} &=\beta_y +u\beta~.  \cr} \eqno{(39)}$$
{}From the first equation of (31), we find that $\alpha,\beta$ evolve via
$$\eqalign{
\alpha_t&={\textstyle{1\over 4}}\alpha u_x
         -{\textstyle{1\over 2}}\alpha_x u
         -{\textstyle{3\over 4}}\alpha \partial_x^{-1}u_{y}
         +\alpha_{xy} ~,\cr
\beta_t&={\textstyle{1\over 4}}\beta u_x
         -{\textstyle{1\over 2}}\beta_x u
         -{\textstyle{3\over 4}}\beta \partial_x^{-1}u_{y}
         +\beta_{xy}~. \cr }\eqno{(40)}$$
The evolutions (10) and (40) of course preserve the relations (39).
Setting $q=\beta/\alpha$ (assuming $\alpha$ nonzero) we obtain
$$ {{\alpha_x}\over {\alpha}}= {{q_y-q_{xx}}\over{2q_x}}
         \eqno{(41)}$$
{}from which
$$ u=-{\textstyle{1\over 2}}\left({{q_{xxx}}\over{q_x}}-{{3q_{xx}^2}
     \over{2q_x^2}} \right)+\left({{q_y}\over{q_x}}\right)_x
    +{\textstyle{1\over 4}}\left({{q_y}\over{q_x}}\right)^2
-{\textstyle{1\over 2}}\partial_x^{-1}\left({{q_y}\over{q_x}}\right)_y~.
    \eqno{(42)}$$
$q$ satisfies the evolution equation
$$ q_t={\textstyle{1\over 4}}q_x\left(
  \left({{q_{xxx}}\over{q_x}}-{{3q_{xx}^2}\over{2q_x^2}} \right)
 +{\textstyle{3\over 2}} \left({{q_y}\over{q_x}}\right)^2
 +3 \partial_x^{-1}\left({{q_y}\over{q_x}}\right)_y \right)  ~.
    \eqno{(43)}$$
This equation has appeared a number of times in the literature
(see for example [12] and references therein); the novelty of our
approach is that we can write down the infinite
dimensional symmetry which leaves the evolution (43) and
$u$ as given in (42) invariant.
The first of the symmetries in (37) gives the obvious M\"obius
symmetry
$$ q\rightarrow {{aq+b}\over{cq+d}}\qquad\qquad (a,b,c,d ~{\rm constant}).
          \eqno{(44)}$$
The second symmetry in (37) becomes
$$ q\rightarrow {{yq+R}\over{y+S}}~,  \eqno{(45)}$$
where $R,S$ (defined by $\alpha R=(g_1)_{12}$,
$\alpha S=(g_1)_{11}$) are functions satisfying
$$ \eqalign{
R_y&=R_{xx}-q-R_x\left({{q_{xx}-q_y}\over{q_x}}\right) \cr
S_y&=S_{xx}-1-S_x\left({{q_{xx}-q_y}\over{q_x}}\right) \cr
       }\eqno{(46)}$$
$$ \eqalign{
R_t&=R_{xxx}-{\textstyle{3\over 2}}R_{xx}
     \left({{q_{xx}-q_y}\over{q_x}}\right)
    -{\textstyle{3\over 4}}R_x\left(
       {{q_{xxx}}\over{q_x}}-{{3q_{xx}^2}
     \over{2q_x^2}}+{{2q_yq_{xx}}\over{q_x^2}}
    -{\textstyle{1\over 2}}\left({{q_y}\over{q_x}}\right)^2
     -\partial_x^{-1}\left({{q_y}\over{q_x}}\right)_y
    \right)\cr
S_t&=S_{xxx}-{\textstyle{3\over 2}}S_{xx}
     \left({{q_{xx}-q_y}\over{q_x}}\right)
    -{\textstyle{3\over 4}}S_x\left(
       {{q_{xxx}}\over{q_x}}-{{3q_{xx}^2}
     \over{2q_x^2}} +{{2q_yq_{xx}}\over{q_x^2}}
    -{\textstyle{1\over 2}}\left({{q_y}\over{q_x}}\right)^2
     -\partial_x^{-1}\left({{q_y}\over{q_x}}\right)_y
   \right)   .\cr }\eqno{(47)}$$
Formula (42) (or rather the formula for $u_x$ obtained from (42),
thereby eliminating the awkward integration symbol on the right hand
side) thus gives a remarkable infinite dimensional
extension of the standard Schwartzian derivative, with $u_x$
invariant under the symmetries generated by (44) and (45),
where $R,S$ satisfy (46).
In fact it can be shown that the map (44) is a special case of (45).
It would be interesting, but probably
rather hard, to prove the uniqueness of this invariant, in the
sense of [2]. For the reader concerned by the asymmetry between
$R$ and $S$ in (46), we note that (46) and (47) imply identical
$y$ and $t$ evolutions for the two functions $R+yq$ and $S+y$,
and actually both of these functions satisfy the UrKP equation (43).

Two tasks remain: to understand the place of the MKP equation (9) in
our framework, and to obtain B\"acklund transformations. Above, we have
defined a projective action of the ring $\cal H$ (defined in (33))
on $q$, and we expect the MKP field to be a combination of $q$ and its
derivatives, invariant under some subring of
$\cal H$ \footnote{$^3$}{While  $\cal H$ has a ring structure,
as do the subsets of $\cal H$ that will interest us,
the additive structure is unimportant for us, so it
might be clearer for some to replace the words ``ring'' and
``subring'' here and in what follows by the words
``monoid'' and ``submonoid''.}. Apparently
a number of candidate subrings are available, both infinite
dimensional (e.g. $\{h\vert {\rm all}~{h_n}~{\rm upper~
triangular}\}$), and finite dimensional (e.g. $\{h\vert h_n=0,n>0,
h_0~{\rm constant}\}$),
but on reflection the necessary subring must be obtained from
$\cal H$ by placing restrictions only on $h_0$.
At present we do not have a general procedure for constructing an
invariant corresponding to any given subring. The one tool we do have
for constructing invariants exploits the notion of {\it gauge
symmetry}, which is the invariance of equations (30) and (31) under
transformations
$$ \eqalign{ g&\rightarrow{\bar{g}}=sg\cr
             M&\rightarrow{\bar{M}}=sMs^{-1}-s_xs^{-1}\cr
             P&\rightarrow{\bar{P}}=sPs^{-1}-s_ts^{-1}.\cr
             }\eqno{(48)}$$
Here
$$ s=\pmatrix{ s_1(x,y,t)&0\cr s_{\rm op}
           &s_2(x,y,t)\cr}~,    \eqno{(49)}$$
where $s_{\rm op}$ is a finite order differential operator in $y$ (with
coefficients functions of $x,y,t$), and $s_1,s_2$ are functions.
This choice of $s$ allows
$s^{-1}$ and all the necessary multiplications in (48) to be defined.
One can exploit gauge symmetry to bring $g$ to a normal
form; but then $\bar{M},\bar{P}$ will not be invariant under all the
symmetries (33), since although $M,P$ are, the gauge transformation
$s$ needed to bring $g$ to this normal form is not. However it is
reasonable to hope that in fact $\bar{M},\bar{P}$ will be
invariant under some subring of $\cal H$ (c.f. [7]).

As an example of such a procedure we consider gauge transformations
that bring $g$ to a form where $(g_0)_{21}=0$.
Any  gauge transformation satisfying
$$ s_{\rm op}\vert \alpha = s_2\alpha_x     \eqno{(50)}$$
does this,
where $s_{\rm op}\vert\alpha$ denotes the function obtained by letting
$s_{\rm op}$ act on $\alpha$. One way to satisfy (50) is to take
$s_1=s_2=1$, $s_{\rm op}=\alpha_x/\alpha$. Then writing
$j=-2\alpha_x/\alpha=(q_{xx}-q_y)/q_x$ we find
$$ \eqalign{ \bar{M} & =
\pmatrix{{\textstyle{1\over 2}}j&1\cr
   \noalign{\smallskip}
\partial_y+{\textstyle{1\over 2}}\partial_x^{-1}j_y
        &-{\textstyle{1\over 2}}j} \cr
   \noalign{\smallskip}
\bar{P} & =
\pmatrix{ {\textstyle{1\over 2}}j\partial_y + {\textstyle{1\over 8}}f_1
        & \partial_y + {\textstyle{1\over 4}}f_2 \cr
   \noalign{\smallskip}
\partial_y^2-{\textstyle{1\over 4}}f_3\partial_y - {\textstyle{1\over 8}}f_4
    &    -{\textstyle{1\over 2}}j\partial_y-{\textstyle{1\over 8}}f_5
    \cr}, \cr}
\eqno{(51)}$$
$$ \eqalign{ f_1&=j_{xx}-{\textstyle{1\over 2}}j^3-j\partial_x^{-1}j_y
                   +3\partial_x^{-1}(jj_y)+3\partial_x^{-2}j_{yy} \cr
  f_2&=j_x-{\textstyle{1\over 2}}j^2-\partial_x^{-1}j_y \cr
  f_3&=j_x+{\textstyle{1\over 2}}j^2-\partial_x^{-1}j_y \cr
  f_4&=  (\partial_x^{-1}j_y)^2
         +(j_x+{\textstyle{1\over 2}}j^2)\partial_x^{-1}j_y
         -4\partial_x^{-1}j_{yy}\cr
  f_5&=j_{xx}-{\textstyle{1\over 2}}j^3+2j_y-j\partial_x^{-1}j_y
                   -3\partial_x^{-1}(jj_y)-3\partial_x^{-2}j_{yy}.
         \cr }$$
The equation of zero curvature for $\bar{M},\bar{P}$ gives
the MKP equation (9)\footnote{$^4$}{Note this zero curvature form for
MKP is not canonical (in the sense of Drinfeld and Sokolov),
unlike its counterpart for MKdV found by
replacing $\partial_y$ by $\lambda$ and setting
$j_y$ to zero in $\bar{P}$ and $\bar{M}$.}.
The invariance of $j$ is easily
found to be the subring  of elements of $\cal H$ with
$(h_0)_{11}$ constant and $(h_0)_{21}$ zero.
This is a maximal infinite-dimensional subring of $\cal H$,
the action of which on $q$ is generated by the transformations
$$ q\rightarrow yq+R~,   \eqno{(52)}$$
which include M\"obius transformations (44) with $c=0$ as a special case.
Transformations (52) are a special case of transformations
(45) corresponding to the choice $S=1-y$. Under the general
transformation (45) one finds
$$ j\rightarrow j-{{2S_x}\over{S+y}} ~. \eqno{(53)}$$
We will use this to find a B\"acklund transformation shortly.
But first we complete this section on the origin of the MKP
equation by noting that the KP-Miura map can be
obtained from the definition of $j$ and the first equation
of (39), and expresses the fact that any invariant under
$\cal H$ is necessarily an invariant under the subring just given.

In this paper we will
only consider the simplest B\"acklund transformations for
MKP and KP, deferring a more detailed study, and comparison with the
results of [12],[13],[14] for [9]. From above we have a symmetry
of MKP given by equation (53), where $S$ satisfies
$ S_y=S_{xx}-1-S_xj$. Similarly, the simpler transformation (44)
with $a=d=0$, $b=c=1$ gives a symmetry
$$ j\rightarrow j- {{2q_x}\over q}~.   \eqno{(54)}$$
These symmetries are equivalent, and can be summarized
as the B\"acklund transformation
$$\eqalign{ j&\rightarrow j-2q_x/q \cr
            q_y&=q_{xx}-jq_x \cr
            q_t&=q_{xxx}-{\textstyle{3\over 2}}jq_{xx} -
                 {\textstyle{3\over 4}}(j_x-{\textstyle{1\over 2}}
                  j^2+\partial_x^{-1}j_y)q_x~. \cr}
                                             \eqno{(55)}$$
One can directly check that this is a strong B\"acklund
transformation, that the composite of two such
transformations is of the same form, and that such transformations
leave $u$ invariant, as expected. (55) becomes
powerful when used in tandem with the more obvious
B\"acklund transformation of MKP, namely $j\rightarrow-j$,
$y\rightarrow-y$. This latter transformation does not leave $u$
invariant, but rather induces the well known linear strong
B\"acklund transformation of KP [13]:
$$ \eqalign{u&\rightarrow u-2(\alpha_x/\alpha)_x\cr
            \alpha_y&=\alpha_{xx}-u\alpha \cr
\alpha_t&=  \alpha_{xxx}
         -{\textstyle{3\over 2}}u\alpha_x
         -{\textstyle{3\over 4}}(u_x+\partial_x^{-1}u_{y})\alpha
               \cr } \eqno{(56)}$$
(note KP is invariant under $y\rightarrow -y$).  Now since the
square of this ``obvious'' transformation for MKP is the identity,
it is not apparent that one can use it to find more than one
solution {}from a given one; but using the transformation (55) as
well allows one (at least in principle) to find chains of solutions
of MKP (and thus KP). The combination of applying the obvious
transformation and then a transformation of the form (55) is
equivalent to one of the two fundamental gauge transformations of
[13]; the second arises from consideration of the system (32).

One other B\"acklund transformation has essentially already appeared
in this paper, and therefore merits a mention: the transformation
(52), with $R$ satisfying the first equations of (46) and (47) is
a strong B\"acklund transformation for the UrKP equation (43).

\vskip .2in

\noindent{\bf Concluding Remarks}

\smallskip

As we have mentioned, we intend in [9] to give a fuller account of
the various different notions of ``UrKP'' that can be found, and of the
associated symmetries, modified equations and B\"acklund transformations.
Many other issues remain to be addressed, such as setting our results
on UrKP in a hamiltonian framework (c.f.[1]) and  understanding the
relationship of our formulation of MKP with the existing formulations
(see [12], [14], [15], [16]). There also remain certain
open questions in the UrKdV formalism - it is not yet clear whether
Wilson's ideas can be used to understand the existence of all the
equations related by Miura maps to KdV (see [5]), and it would also
be interesting to see if Wilson's ideas give us insight into the
B\"acklund transformations of KdV and MKdV that we have not discussed,
namely those with dependence on dimensionful parameters. Possibly
the most important direction for further research though involves
the {\it physical application} of Wilson's ideas. The upshot of
Wilson's ideas for KdV and their extension for KP is that whenever
a modified KdV or KP equation appears in a physical context, then
there is a hidden symmetry waiting to be unearthed. This last
direction is currently being actively pursued.

\vskip.2in

\noindent{\bf Acknowledgements}

\smallskip

D.A.D. is supported in part by the Natural Science and Engineering
Research Council of Canada, and by the Fonds F.C.A.R. du Qu\'ebec.
Most of this work was done while J.S. was at the Institute for
Advanced Study, Princeton, NJ, supported by a
grant in aid from the U.S.Department of Energy, \#DE-FG02-90ER40542.
J.S. is currently supported by a Guastella fellowship of the Rashi
Foundation.

\vskip .2in

\noindent{\bf References}

\noindent
\item{[1]} G.Wilson, {\it Phys.Lett.A} {\bf 132} (1988) 45.
\item{[2]} M.Lavie, {\it Can.J.Math.} {\bf 21} (1969) 235, section 3.
\item{[3]} J.Schiff,
talk given at the NSERC-CAP Workshop ``Quantum Groups, Integrable
Models and Statistical Systems'', Kingston, Ontario (July 1992).
Written version to be published in workshop proceedings.
\item{[4]} B.G.Konopelchenko,{\it Phys.Lett.A} {\bf 92} (1982) 323.
\item{[5]} S.I.Svinolupov,V.V.Sokolov and R.I.Yamilov,
    {\it Soviet Math.Dokl.} {\bf 28} (1983) 165.
\item{[6]} R.Sasaki and I.Yamanaka, {\it Adv.Stud.Pur.Math.}
    {\bf 16} (1988) 271.
\item{[7]} I.McIntosh, {\it Proc.Roy.Soc.Edinburgh} {\bf 115A} (1990)
       367.
\item{[8]} P.M.Santini and A.S.Fokas, {\it Comm.Math.Phys.}
     {\bf 115} (1988) 115.
\item{[9]} D.A.Depireux and J.Schiff, in preparation.
\item{[10]} I.Bakas and E.B.Kiritsis, {\it Prog.Theo.Phys.Suppl.}
    {\bf 102} (1990) 15.
\item{[11]} S.Odake and T.Sano, {\it Phys.Lett.B} {\bf 258} (1991) 369.
\item{[12]} W.Oevel and C.Rogers,
   {\it Rev.Math.Phys.} {\bf 5} (1993) 299.
\item{[13]} L.-L.Chau,J.C.Shaw and H.C.Yen,
    {\it Comm.Math.Phys.} {\bf 149} (1992) 263.
\item{[14]} B.G.Konopelchenko and W.Oevel,
   {\it Publ.R.I.M.S.Kyoto} {\bf 29} (1993) 581.
\item{[15]} M.Jimbo and T.Miwa, {\it Publ.R.I.M.S.Kyoto},
    {\bf 19} (1983) 943.
\item{[16]} V.G.Kac and D.H.Peterson, in {\it Syst\`emes
    Dynamiques Non Lin\'eaires: Int\'egrabilit\'e et
    Comportement Qualitatif}, ed. P.Winternitz (1986), Les
    Presses de l'Universit\'e de Montr\'eal.

\bye